\documentclass[aps,prl,twocolumn,showpacs,groupedaddress]{revtex4}  
\usepackage{graphicx}  
\usepackage{dcolumn}   
\usepackage{bm}        
\usepackage{amssymb}   
\usepackage{xspace}

\begin{document}
\hspace{5.2in} \mbox{Fermilab-Pub-04/08-010-E}

\newcommand{\dzero}     {D0}
\newcommand{\ttbar}     {\mbox{$t\bar{t}$}\xspace}
\newcommand{\bbbar}     {\mbox{$b\bar{b}$}\xspace}
\newcommand{\ccbar}     {\mbox{$c\bar{c}$}\xspace}
\newcommand{\ppbar}     {\mbox{$p\bar{p}$}\xspace}
\newcommand{\pythia}    {\sc{pythia}}
\newcommand{\alpgen}    {\sc{alpgen}}
\newcommand{\qq}        {\sc{qq}}
\newcommand{\evtgen}    {\sc{evtgen}}
\newcommand{\tauola}    {\sc{tauola}}
\newcommand{\geant}     {\sc{geant}}
\newcommand{\singletop}   {\sc{singletop}}
\newcommand{\met}       {\mbox{$\not\!\!E_T$}\xspace}
\newcommand{\metcal}    {\mbox{$\not\!\!E_{Tcal}$}\xspace}
\newcommand{\rar}       {\rightarrow}
\newcommand{\eps}       {\epsilon}

\newcommand{\dilepton}  {$t\overline{t}_{dilepton}$\xspace}
\newcommand{\ljets}     {\mbox{$\ell$+jets}\xspace}
\newcommand{\ejets}     {\mbox{$e$+jets}\xspace}
\newcommand{\mujets}    {\mbox{$\mu$+jets}\xspace}
\newcommand{\Zmumu}     {$Z\to\mu\mu$\xspace}
\newcommand{\Zee}       {$Z\to ee$\xspace}
\newcommand{\wplus}     {$W$+jets\xspace}
\newcommand{\wplusfour} {$W$+4jets\xspace}

\title{\boldmath Simultaneous measurement of the ratio
$  {\cal {B}}(  t\rightarrow Wb )/{\cal{B}}(t\rightarrow Wq )$ and the top quark pair 
production cross section with the \dzero~detector at $ \sqrt{s}=1.96$~TeV}
%
\author{V.M.~Abazov$^{36}$}
\author{B.~Abbott$^{76}$}
\author{M.~Abolins$^{66}$}
\author{B.S.~Acharya$^{29}$}
\author{M.~Adams$^{52}$}
\author{T.~Adams$^{50}$}
\author{E.~Aguilo$^{6}$}
\author{S.H.~Ahn$^{31}$}
\author{M.~Ahsan$^{60}$}
\author{G.D.~Alexeev$^{36}$}
% ...

\affiliation{\vspace{0.1 in}(The D\O\ Collaboration)\vspace{0.1 in}}
% ...
\affiliation{$^{6}$University of Alberta, Edmonton, Alberta, Canada,
                Simon Fraser University, Burnaby, British Columbia, Canada,
                York University, Toronto, Ontario, Canada, and
                McGill University, Montreal, Quebec, Canada}
% ...
\affiliation{$^{29}$Tata Institute of Fundamental Research, Mumbai, India}
% ...
\affiliation{$^{31}$Korea Detector Laboratory, Korea University, Seoul, Korea}
% ...
\affiliation{$^{36}$Joint Institute for Nuclear Research, Dubna, Russia}
% ...
\affiliation{$^{50}$Florida State University, Tallahassee, Florida 32306, USA}
% ...
\affiliation{$^{52}$University of Illinois at Chicago,
                Chicago, Illinois 60607, USA}
% ...
\affiliation{$^{60}$Kansas State University, Manhattan, Kansas 66506, USA}
% ...
\affiliation{$^{66}$Michigan State University,
                East Lansing, Michigan 48824, USA}
% ...
\affiliation{$^{76}$University of Oklahoma, Norman, Oklahoma 73019, USA}
% ...

\date{January 8, 2008}

\begin{abstract}
We present the first simultaneous measurement of the ratio of branching 
fractions, $R={\cal{B}}(t \rightarrow Wb)/{\cal{B}}(t \rightarrow Wq)$, with $q$ 
being a $d$, $s$, or $b$ quark, and the top quark pair production cross 
section $\sigma_{t\bar{t}}$ in the lepton plus jets channel 
using $0.9~\text{fb}^{-1}$ of $p\bar{p}$ collision data at $\sqrt{s}=1.96$~TeV
collected with the D0 detector. 
We extract $R$ and $\sigma_{t\bar{t}}$ by analyzing samples of 
events with 0, 1 and $\ge 2$ identified $b$ jets. 
We measure $R = 0.97^{+0.09 }_{-0.08}\text{~(stat+syst)}$ 
and $\sigma_{t\bar{t}} = 8.18^{+0.90}_{-0.84}\text{~(stat+syst)} \pm
0.50\text{~(lumi)}~\text{pb}$, in agreement with the standard model prediction.

\end{abstract}

\maketitle

Within the standard model (SM) the top quark decays to a $W$ boson 
and a down-type quark $q$ 
($q$ = $d, s, b$) with a rate proportional to the squared 
Cabibbo-Kobayashi-Maskawa (CKM) matrix element, $|V_{tq}|^2$ ~\cite{CKM}. 
Under the assumption of three  
fermion families and a unitary $3\times3$ CKM matrix, the $|V_{tq}|$ elements are 
severely constrained: 
$|V_{td}|=(7.4 \pm 0.8)\cdot 10^{-3}$, 
$|V_{ts}|=(40.6\pm2.7)\cdot 10^{-3}$ and $|V_{tb}|=0.999100^{ +0.000034}_{ -0.000004}$~\cite{pdg}.
However, in several extensions of the SM 
the $3\times 3$ 
CKM submatrix would not appear unitary and $|V_{tq}|$ elements can 
significantly deviate from their SM values. This would affect the rate for  
single top quarks production via the electroweak interaction
~\cite{stopPRL} and 
the ratio $R$ of the top quark branching fractions, which 
can be expressed in terms of the CKM matrix elements as 
\begin{eqnarray*}
\label{eq:Rdef}
R = \frac{{ \cal B}(t \rightarrow Wb)}{{ \cal B}(t \rightarrow Wq)} & = &
\frac{\mid V_{tb}\mid^2}{\mid V_{tb}\mid^2 + \mid V_{ts}\mid^2 + \mid V_{td}\mid^2}  \;.
\end{eqnarray*}
A precise measurement of $R$ is therefore a necessary ingredient for 
performing direct measurements, free of assumptions about the
number of quark families or the unitarity of the CKM matrix,  
of the $|V_{tq}|$ elements  
via the combination with future measurements of the  
single top quark production in $s$ and $t$ channels \cite{Vtb_theory}.  

In this Letter, we report the first simultaneous measurement of $R$ and the top 
quark pair ($t\bar{t}$) production cross section $\sigma_{t\bar{t}}$. 
$R$ was measured by the CDF and D0 collaborations~\cite{CDFref,D0BR}.  
The simultaneous measurement of $R$ and $\sigma_{t\bar{t}}$, in contrast to   
previous measurements \cite{p14btag,CDFbtag}, allows one to extract $\sigma_{t\bar{t}}$ 
without assuming ${\cal B}(t \rightarrow Wb)=1$, and to achieve a higher precision 
on both quantities by exploiting their different sensitivity to systematic 
uncertainties. 

The current measurement is based on data collected with the D0 detector 
\cite{run2det} between August 2002 and December 2005 at the Fermilab Tevatron
{\mbox{$p\bar p$}}\ collider at {\mbox{$\sqrt{s}$ =\ 1.96\ TeV}}, 
corresponding to an integrated luminosity of 
about $0.9~\text{fb}^{-1}$, approximately four times 
larger than that of our previous measurement~\cite{D0BR}. 
The analysis uses the top quark pair decay channel 
{\mbox{$ t\overline{t} \rightarrow W^{+} q W^{-} \overline{q}$}}, with the
subsequent decay of one $W$ boson into two quarks, and the other  
one into an electron or muon and a neutrino. This is  
referred to as the lepton plus jets (\ljets) ~channel.  
We select a data sample enriched in $t\bar{t}$ events by  
requiring $\ge 3$ jets with transverse momentum $p_T>20$~GeV 
and pseudorapidity $|\eta|<2.5$~\cite{p14topo}, 
one isolated electron with $p_T>20$~GeV and $|\eta|<1.1$ or muon 
with $p_T>20$~GeV and $|\eta|<2.0$, and missing transverse energy 
$\met>20$~GeV in the \ejets and $\met>25$~GeV in the \mujets~ channel. 
Additionally, the leading jet $p_T$ is required to exceed  
$40$~GeV.
Events containing a second isolated electron or muon with $p_T>15$~GeV are rejected.
The lepton isolation criteria are based on calorimeter and tracking 
information. Details of lepton, jets and $\met$  
identification are described elsewhere~\cite{p14topo}. 

We identify $b$-jets using a
neural-network tagging algorithm~\cite{btagging}. It 
combines variables that characterize the presence and properties of 
secondary vertices and tracks with high impact parameter inside the jet.
In the simulation, we assign a probability for each jet to be $b$-tagged based 
on its flavor, $p_T$, and $\eta$. These probabilities are determined 
from data control samples, and can be combined to yield a probability 
for each $t\bar{t}$ event to have 0, 1, or $\ge 2$ $b$-tagged jets~\cite{p14btag}. 

We split the selected \ljets sample into subsamples according 
to the lepton flavor ($e$ or $\mu$), jet multiplicity (3 or $\ge 4$ jets) and 
number of identified $b$-jets (0, 1 or $\ge 2$), thus obtaining 12 disjoint 
data sets.  
We fit simultaneously $R$ and $\sigma_{t\bar{t}}$ to the observed number of 
1 $b$ tag and $\ge 2$ $b$ tag events, and, in 0 $b$ tag events with 
$\ge 4$ jets, to the shape of a discriminant 
$\cal{D}$ that exploits kinematic differences between the background and the 
$t\bar{t}$ signal and which is described in detail below. As the 
signal-to-background ratio is about five times smaller in events with 0 $b$ tags and three jets we do not use a discriminant for that subsample. 

The dominant background is the production of $W$ bosons in association with 
heavy and light flavor jets ($W$+jets). Smaller contributions arise 
from $Z$+jets, diboson and single top quark production.
Multijet events 
enter the selected sample if a jet is misidentified as an electron  
(\ejets), or a muon coming from either a semileptonic heavy 
quark decay or an in-flight pion or kaon decay in a light flavor jet   
appears isolated (\mujets). 

 
We model the signal and  backgrounds other 
than multijet using a Monte Carlo 
(MC) simulation. The processes $W$+jets and $Z$+jets are generated with 
the {\alpgen}~{\sc 2.05}~\cite{alpgen} leading-order event generator for the
multi-parton  matrix element calculation and {\pythia}~{\sc 6.323}~\cite{pythia} 
for subsequent parton showering and hadronization. Diboson samples are generated with 
{\pythia} and single top quark production is modeled using 
the {\singletop} ~\cite{singletop} 
generator. The \ttbar signal is simulated with {\pythia} for a top quark 
mass of $m_{top}=175$~GeV and includes 
three decay modes $t\bar{t} \rightarrow W^{+}bW^{-}\bar{b}$,  
$t\bar{t}\rightarrow W^{+}bW^{-}\bar{q}_l$ (or $t\bar{t}\rightarrow W^{+}q_lW^{-}\bar{b}$) and 
$t\bar{t} \rightarrow W^{+}q_lW^{-}\bar{q}_l$, 
where $q_l$ denotes a light down-type  ($d$ or $s$) quark. 
These three decay modes are referred to as $bb$, $bq_l$ and $q_lq_l$.
The generated events are processed through 
the {\geant}-based~\cite{geant} simulation of the {\dzero} detector. The same reconstruction 
algorithm as for data is used. 
Additional corrections~\cite{p14topo} are  applied to the   
reconstructed objects to improve the agreement between data and simulation. 
   
The determination of the background composition starts with the 
evaluation of the multijet background for each jet multiplicity and  
lepton flavor before $b$-jet tagging by counting 
events in the corresponding control data samples and applying 
the matrix method~\cite{p14btag}. We estimate the number 
of events with a lepton originating from a
$W$ or $Z$ boson decay by subtracting the estimated multijet 
background from the observed event yield before $b$-tagging.  
We further subtract diboson, single top quark and 
$Z$+jets contributions, 
normalized to   
the next-to-leading order cross sections~\cite{nlomc}. 
The remaining data
events are assumed to come from \ttbar and $W$+jets background. 
In the first step of the  
fitting procedure used to extract $\sigma_{t\bar{t}}$ 
and $R$, we assume the ${t\bar{t}}$ contribution predicted by the 
SM~\cite{SMtheory}. In every subsequent step,  
we iteratively re-determine the expected number of \ttbar events 
and re-evaluate the $W$+jets background.  

\begin{table}[t]
\begin{center}
\caption{Sample composition for the measured $\sigma_{t\bar{t}}$ 
and $R=1$. The uncertainties are statistical only. \label{tab:yields} }
\begin{tabular}{l  l  r@{ $\pm$ }l  r@{ $\pm$ }l  r@{ $\pm$ }l }
\hline
      \hline
 $N_{\text{jets}}$   & sample & \multicolumn{2}{c}{0 $b$ tags} &
 \multicolumn{2}{c}{1 $b$ tag } & \multicolumn{2}{c}{$\ge$ 2 $b$ tags} \\
\hline 

3 & $W$+jets   & $1394.4 $ & $48.4$  &  $102.5 $ & $ 3.5$ & $8.3 $ & $ 0.3$ \\
& Multijet   & $287.4  $ & $35.9$   & $28.1 $ & $ 3.5$   & $3.3$ & $ 0.4$ \\
& Other      & $254.0 $ & $1.9$    & $29.4 $ & $ 0.2$   & $5.2 $ & $ 0.4$ \\
& $t\bar{t}$ & $ 109.7 $ & $0.4$   &  $143.3$ & $ 0.5$ & $54.3 $ & $ 0.2$\\& Total      & $2045.5  $ & $60.3$  & $303.3 $ & $ 5.0$  & $71.2 $ & $ 0.5$ \\
& Observed & \multicolumn{2}{c}{2050} &                \multicolumn{2}{c}{294}               &  \multicolumn{2}{c}{76} \\ 
\hline
4 & $W$+jets & $188.2 $ & $15.0$ &  $17.3 $ & $ 1.4$ & $1.8 $ & $ 0.1$ \\
& Multijet & $66.9 $ & $9.9$ &   $6.6 $ & $ 1.0$  & $0.8 $ & $ 0.1$ \\
& Other & $62.2  $ & $0.9$ &      $8.0 $ & $ 0.1$  & $1.7 $ & $ 0.0$ \\
& $t\bar{t}$ & $83.8  $ & $0.3$ & $126.4 $ & $ 0.5$   & $64.2 $ & $ 0.2$ \\
& Total & $401.1 $ & $18.0 $ &   $158.3 $ & $ 1.8$ & $69.5 $ & $ 0.3$ \\
& Observed & \multicolumn{2}{c}{389} &                 \multicolumn{2}{c}{179}               &  \multicolumn{2}{c}{58} \\
\hline
      \hline
\end{tabular}
\end{center}
\end{table}

\begin{figure*}[ht]
\begin{center}
\setlength{\unitlength}{1.0cm}
\begin{picture}(18.0,5.0)
\put(0.2,0.2){\includegraphics[width=5.0cm]{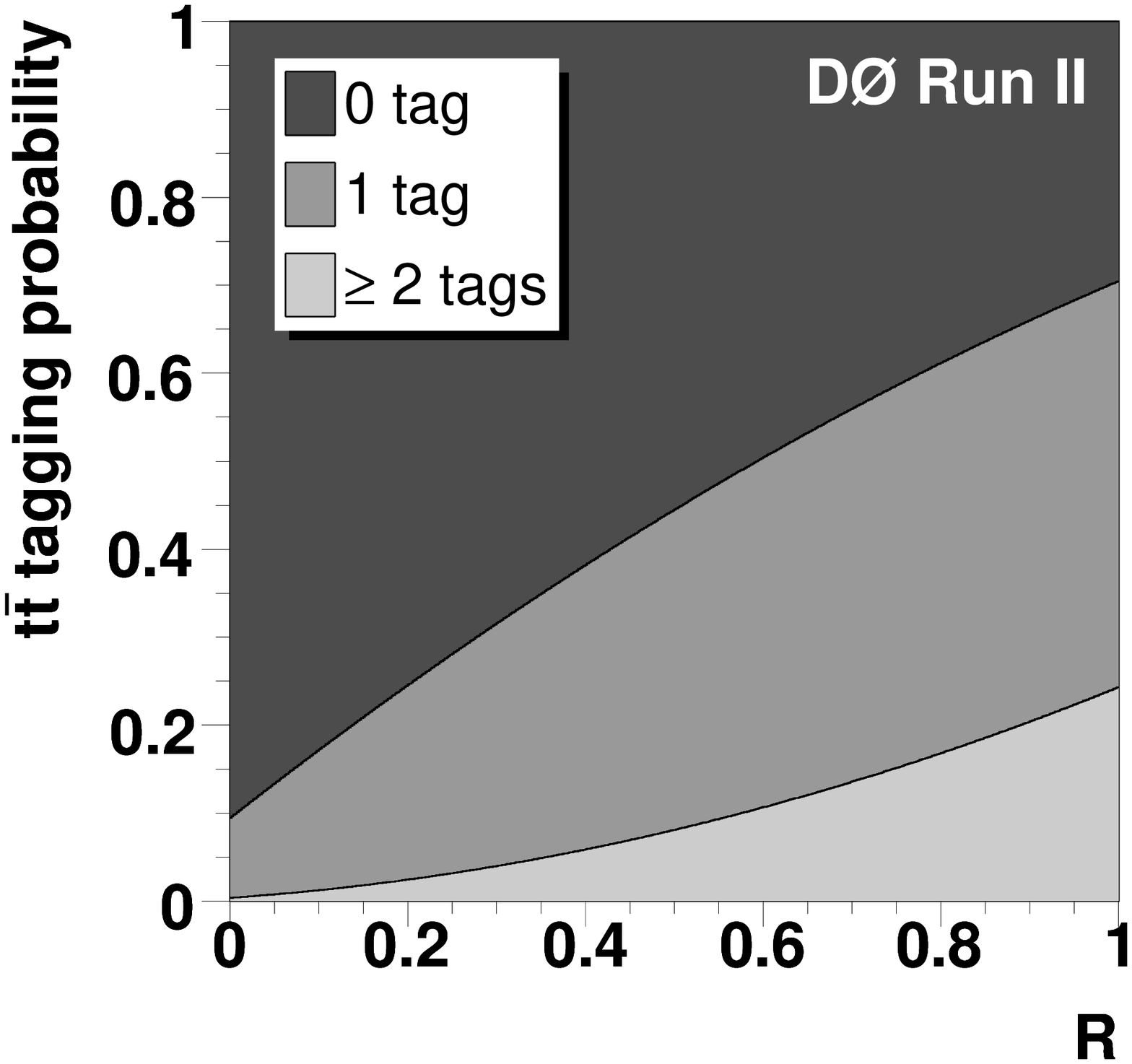}  }
\put(6.2,0.2){\includegraphics[width=5.0cm]{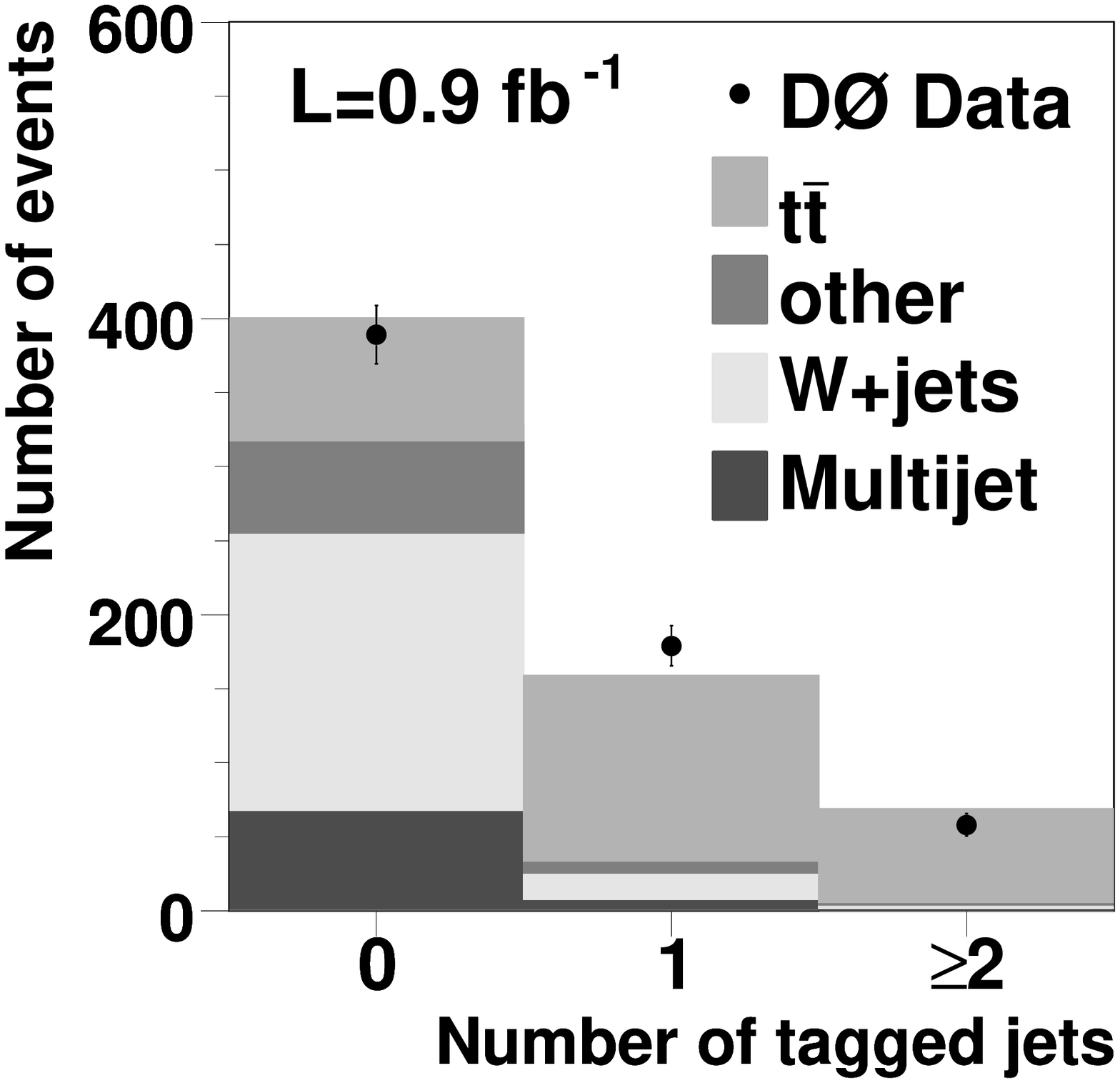} }
\put(12.2,0.2){\includegraphics[width=5.0cm]{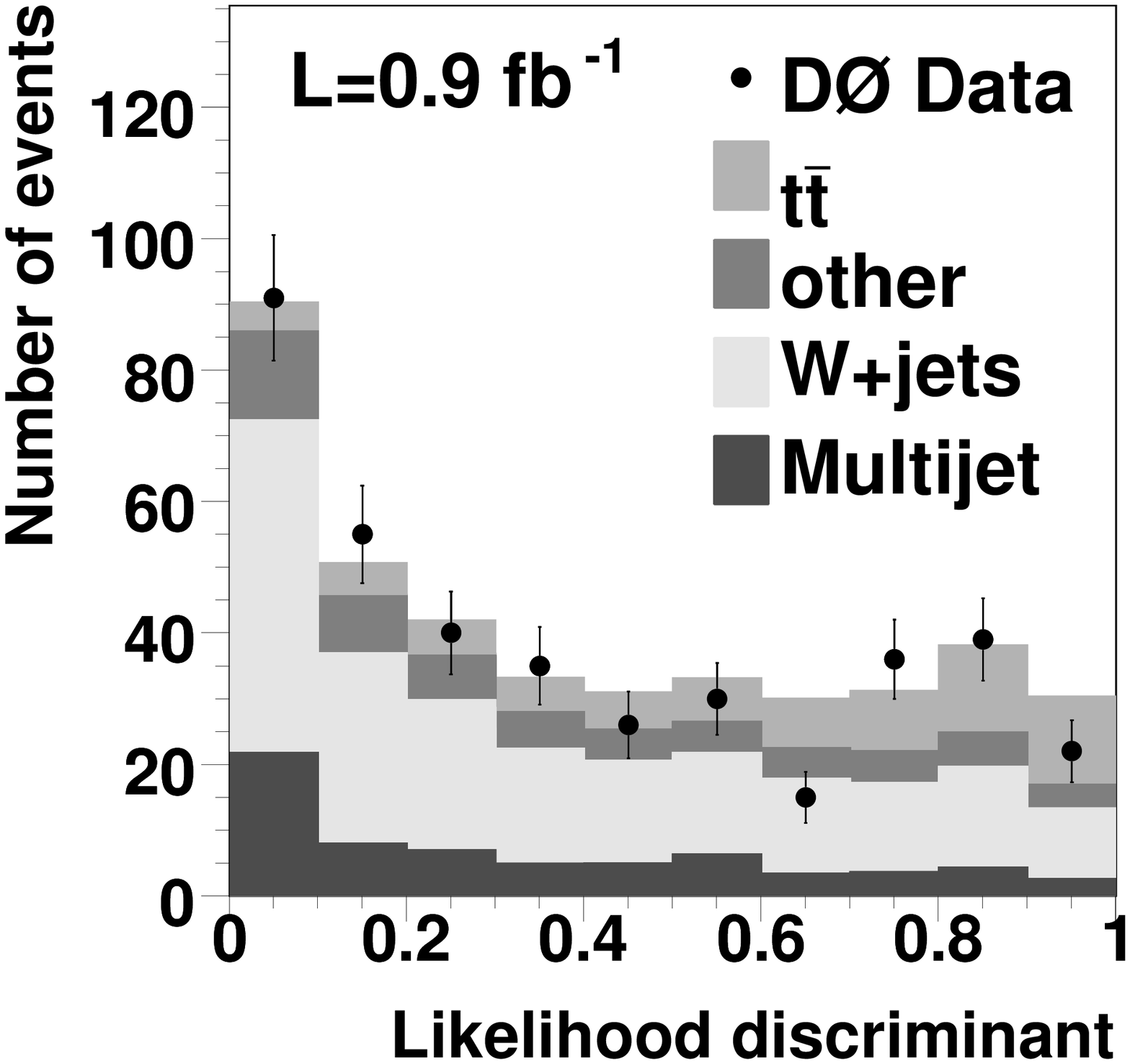} }
\put(3.0,5.0){(a)}
\put(9.0,5.0){(b)}
\put(15.0,5.0){(c)}
\end{picture}
\caption{\label{fig:tagbins_plots_medium_pre} (a) Fractions of events with 
0, 1 and $\ge 2$ $b$ tags as a function 
of $R$ for $t\bar{t}$~ events with $\ge 4$ jets; 
(b) predicted and observed number of events in  
the 0, 1 and $\ge 2$ $b$ tag samples for the measured $R$ and $\sigma_{t\bar{t}}$ 
for events with $\ge 4$ jets and (c) predicted and observed discriminant
distribution in the 0 $b$ tag sample with $\ge 4$ jets.}
\end{center}
\end{figure*}

Since the probability to tag a \ttbar~ event depends on the flavor of the jets, 
it depends on $R$. 
We estimate the acceptance and tagging probabilities 
for each of the three \ttbar decay modes $bb$, $bq_l$ and $q_lq_l$. 
Thus, the probability for a $t\bar{t}$ event to pass our selection criteria 
and to have $n$ $b$-tagged jets is:  
\begin{eqnarray*}
P_{total}^{n}(t\bar{t})& = & R^2 A(bb)P_t^{n}(bb) + 2R(1-R)A(bq_l) P_t^{n}(bq_l) \nonumber \\
&+& (1-R)^2 A(q_lq_l) P_t^{n}(q_lq_l),
\end{eqnarray*}
where $A$ ($P_t^{n}$) describes the acceptance (tagging probability) 
for each $t\bar{t}$ decay mode.  
Figure~\ref{fig:tagbins_plots_medium_pre}(a) shows $P_t^{n}$ as a function of $R$
for \ttbar~ events with $\ge 4$ jets and 0, 1 and 
$\ge 2$ $b$ tags. 
Table~\ref{tab:yields} presents the composition of the selected sample 
for the measured $\sigma_{t\bar{t}}$ and $R=1$. 


The topological discriminant $\cal D$~\cite{p14topo} exploits  
the kinematic differences between $t\bar{t}$ and 
$W$+jets events to achieve a better constraint 
on the number of $t\bar{t}$ events in the subsample 
with $\ge 4$ jets and 0 $b$ tags. 
We select variables well-described 
by the background model in 
samples of events with one or two jets that provide a good separation
between signal and $W$+jets background. The optimal set of variables 
is chosen to minimize the expected 
statistical uncertainty on the fitted fraction of $t\bar{t}$ events.  
Due to the differences in acceptance 
and sample composition, the discriminants are constructed from different 
sets of variables in the \ejets and \mujets~channels.    
In the \ejets channel we use five variables: 
 the leading jet $p_T$,
 the maximum $\Delta \cal{R}$~\cite{p14topo} between two of the four leading 
jets,  ${\mathcal A}$, 
 $\mathcal C_M$, and $\mathcal D_M$~\cite{tensor}. 
In the \mujets~channel the discriminant is built from six variables: 
  ${\mathcal A}$,  $\mathcal D_M$,
 the scalar sum of the $p_T$ of the four leading
jets and the muon, 
 the scalar sum of the $p_T$ of
the third and fourth jet in the event,  
 the transverse mass of the vector sum of all jets, and
 the ratio of the mass of the three
leading jets to the mass of the event, defined as the invariant mass 
of the vector sum of all four jets, the lepton from the $W$ decay and the 
missing transverse energy coming from the neutrino. 
The sensitivity to soft radiation and to the underlying event is 
reduced by using only the four highest-$p_T$ jets for the kinematic variables.

The discriminant function is built using simulated 
$W$+jets and \ttbar events.  
We evaluate it for each physics process considered in the analysis and
build corresponding template distributions consisting of ten bins.
For $t\bar{t}$ we obtain a distribution for each of the three decay modes
($bb$, $bq_l$ and $q_lq_l$).  
The shapes of the discriminant distributions for $Z$+jets, diboson and single
top backgrounds are found to be similar to that of the $W$+jets events 
and we use the latter to model them. 
We use a sample of data events selected by requiring that 
the lepton fail the isolation criteria to obtain the discriminant 
shape for the multijet background.

We define a likelihood function as the product of Poisson probabilities 
over all 30 subsamples and bins of the discriminant, where in each   
subsample the expected number of events is estimated as a function
of $R$ and $\sigma_{t\bar{t}}$. We include 12 additional Poisson terms to 
constrain the multijet background prediction in each subsample.
The systematic uncertainties are incorporated in the
fit using nuisance parameters~\cite{nuisance}, each 
represented by a Gaussian term in the binned likelihood. 
In this approach, each source of systematic uncertainty is allowed to 
affect the central value of $R$ and $\sigma_{t\bar{t}}$ during the 
likelihood maximization  procedure, yielding a combined statistical and systematic uncertainty.  

The result of the maximum likelihood fit is:  
\begin{eqnarray*}
R &=&0.97^{+0.09 }_{-0.08}~\text{ (stat+syst)~ and} \\
\sigma_{t\bar{t}} &=& 8.18^{+0.90}_{-0.84}~\text{ (stat+syst)~} \pm 0.50 \;
\text{~(lumi)~pb} \;,
\end{eqnarray*} 
for a top quark mass of $175$~GeV. 
Figure~\ref{fig:tagbins_plots_medium_pre}(b,c) compares the distribution of 
the data to the sum of predicted background and measured $t\bar{t}$ signal 
for $R=0.97$. 
We observe no significant dependence of $R$ on $m_{top}$
within $\pm 10$~GeV around the assumed value while $\sigma_{t\bar{t}}$ 
changes by $\mp 0.09\;\rm pb$ per 1 GeV within the same range.  
We find a correlation between $R$ and $\sigma_{t\bar{t}}$ of -58\%.  
Table \ref{tab:systematics} summarizes 
the statistical and leading systematic uncertainties on $R$ and $\sigma_{t\bar{t}}$ 
excluding the 6.1\% uncertainty on the integrated luminosity~\cite{lumi}.
The contribution of each individual source of uncertainty 
is estimated by fixing all but the corresponding 
Gaussian term in the fit.  
The statistical uncertainty is obtained from the fit with all Gaussian terms fixed.   


The total uncertainty on $R$ is about 9\%, compared to 17\% achieved in the 
previous measurement~\cite{D0BR}. 
The largest uncertainty 
comes from the limited statistics. 
Since the $b$-tagging efficiency drives the distribution of the 
events among the 0, 1 and 2 $b$-tag  
subsamples and is strongly anti-correlated with $R$, the systematic 
uncertainty is dominated 
by the $b$-tagging efficiency estimation, responsible for $\sim$90\% of the 
total systematic uncertainty. 
%

The total uncertainty on $\sigma_{t\bar{t}}$, excluding luminosity, is $\sim$10.5\%, 
representing a 30\% improvement over the previous measurement~\cite{p14btag} 
performed under the assumption of $R=1$. In the latter, the primary 4.7\% relative  
uncertainty comes from the $b$-tagging efficiency estimation while 
in the current measurement it is reduced to 1.2\% because $\sigma_{t\bar{t}}$   
is much less sensitive to the variations of the $b$-tagging efficiency than $R$, 
and the two-dimensional fit takes advantage of this feature. 

\begin{table}[t]
  \begin{center}
    \caption{\label{tab:systematics}{ Summary of uncertainties 
  on $\sigma_{t\bar{t}}$ and $R$. }} 
    \begin{tabular}{lcc}
      \hline
      \hline
      Source  & $\Delta \sigma_{t\bar{t}}$ (pb)  & $\Delta R$ \\
      \hline 
      Statistical                                 &  +0.67 $-$0.64  &  +0.067    $-$0.065  \\
      Lepton identification                       &  +0.32 $-$0.27  &   n/a\\
      Jet energy scale                            &  +0.32 $-$0.23  &   n/a\\ 
      $W$+jets background                         &  +0.21 $-$0.23   &   n/a\\
      Multijet background                         &  +0.17 $-$0.17  &  +0.016    $-$0.016  \\
      Signal modeling                             &  +0.12 $-$0.25  &   n/a\\ 
      $b$-tagging efficiency                       &  +0.10 $-$0.09  &  +0.059    $-$0.047 \\
      Other                                       &  +0.24 $-$0.13  &   +0.015    $-$0.014\\
      \hline
      Total uncertainty                           &  +0.90 $-$0.84  &  +0.092    $-$0.083 \\
      \hline
      \hline
    \end{tabular}
  \end{center}
 \end{table}

\begin{figure}[b]
\begin{center}
  \includegraphics[width=8.5cm]{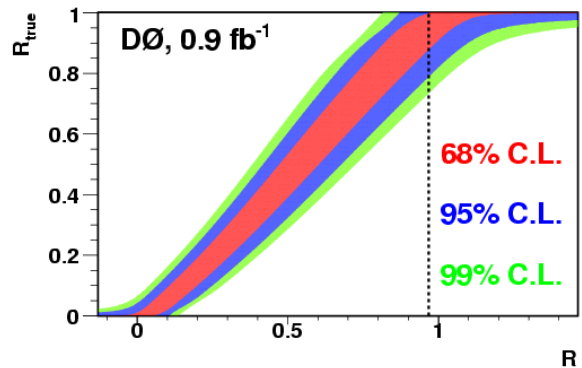}
\end{center}
\vspace{-0.6cm}
    \caption{\label{lowerlimit_FC2D}   The $68\%$ (inner band), $95\%$ (middle band)
    and $99\%$ (outer band) C.L. bands for $R_{\text{true}}$ as a function of $R$. The dotted
    black line indicates the measured value $R=0.97$.}
\end{figure}

We extract a limit on $R$ and $|V_{tb}|$  
following the Feldman-Cousins procedure~\cite{fc_limit}. 
We generate pseudo-experiments with all systematic uncertainties included 
for various input values of $R$ ($R_{\text{true}}$) and apply the 
likelihood-ratio ordering principle. 
We obtain $R > 0.88$ at 68\%~C.L. and $R > 0.79$ at 95\%~C.L.,  
illustrated in Fig.~\ref{lowerlimit_FC2D}. 
From $R$ we determine the ratio of $|V_{tb}|^2$ to the off-diagonal 
matrix elements to be  
$\frac{\mid V_{tb}\mid^2}{\mid V_{ts}\mid^2 + \mid V_{td}\mid^2}>3.8$ at 95\%~C.L. 
Assuming a unitary CKM matrix with three fermion generations we 
derive $|V_{tb}| > 0.89$ at 95\%~C.L.


In summary, we have performed a simultaneous measurement of the 
ratio of branching fractions $R$ and $\sigma_{t\bar{t}}$ yielding 
the most precise measurements to date,  
$R = 0.97^{+0.09 }_{-0.08}~\text{ (stat+syst)}$ and 
$\sigma_{t\bar{t}} = 8.18^{+0.90}_{-0.84}~\text{ (stat+syst)~} \pm 0.50 \;
\text{~(lumi)~pb}$, both in good agreement with the SM. 
This measurement of $R$ will be a key ingredient in a future 
model-independent direct determination of the $|V_{tq}|$ CKM 
matrix elements.





%
We thank the staffs at Fermilab and collaborating institutions, 
and acknowledge support from the 
DOE and NSF (USA);
CEA and CNRS/IN2P3 (France);
FASI, Rosatom and RFBR (Russia);
CAPES, CNPq, FAPERJ, FAPESP and FUNDUNESP (Brazil);
DAE and DST (India);
Colciencias (Colombia);
CONACyT (Mexico);
KRF and KOSEF (Korea);
CONICET and UBACyT (Argentina);
FOM (The Netherlands);
Science and Technology Facilities Council (United Kingdom);
MSMT and GACR (Czech Republic);
CRC Program, CFI, NSERC and WestGrid Project (Canada);
BMBF and DFG (Germany);
SFI (Ireland);
The Swedish Research Council (Sweden);
CAS and CNSF (China);
Alexander von Humboldt Foundation;
and the Marie Curie Program.
%


\begin{thebibliography}{99}
\bibitem{tensor}
The normalized momentum tensor, ${\mathcal M}$, is defined as 
${\mathcal M}_{ij} =  \frac{\Sigma_k p^k_ip^k_j}{\Sigma_k
|\vec{p^k}|^2}$, where $\vec{p^k}$ is the momentum vector of a reconstructed 
object $k$, and $i$ and $j$ are Cartesian coordinates. Aplanarity ${\mathcal A}$, 
${\mathcal C_M}$ and ${\mathcal D_M}$ are 
defined here as $\frac{3}{2}\lambda_{3}$,  
$3 (\lambda_1\lambda_2+\lambda_1\lambda_3+\lambda_2 \lambda_3)$ 
and $27 \lambda_1 \lambda_2 \lambda_3$, respectively,   
where $\lambda_1$, $\lambda_2$ and $\lambda_3$ are the eigenvalues of ${\mathcal M}$.
\end{thebibliography}
\end{document}